\documentclass[aps,prb,twocolumn,floatfix,showpacs,superscriptaddress,longbibliography]{revtex4-2}
\usepackage{amssymb,amsmath,amstext}                
\usepackage{graphicx}                                       
\usepackage{epstopdf}                                       
\usepackage{xcolor}         

\usepackage{amsmath}
\usepackage{bm}                                             
\usepackage{appendix}                                       
\usepackage[utf8]{inputenc}
\usepackage{bbold}
\usepackage{bbm}
\usepackage{ulem}
\usepackage{braket}
\usepackage{latexsym}
\usepackage[colorlinks=true,citecolor=blue,linkcolor=magenta]{hyperref}
\usepackage{etoolbox}  
\usepackage{hyperref}

\def\be{\begin{equation}}
\def\ee{\end{equation}}
\def\bea{\begin{eqnarray}}
\def\eea{\end{eqnarray}}

\def\bi{\begin{itemize}}
\def\ei{\end{itemize}}
\def\ben{\begin{enumerate}}
\def\een{\end{enumerate}}

\newcommand{\mbf}[1]{\mathbf{#1}}
\newcommand{\abs}[1]{\left|#1\right|}

\begin{document}

\title {Spacetime Dynamics of Altermagnetic Magnons}

\author{Ali Emami Kopaei}
\thanks{a.emami-kopaei@uw.edu.pl}
\affiliation{Institute of Theoretical Physics, Faculty of Physics,
University of Warsaw, Pasteura 5, PL-02093 Warsaw, Poland}
\author{Karthik Subramaniam Eswaran}
\affiliation{Szkoła Doktorska Nauk Ścisłych i Przyrodniczych, Wydział Fizyki, Astronomii i Informatyki Stosowanej, Uniwersytet Jagiello\'nski, ulica Profesora Stanisława Łojasiewicza 11, PL-30-348 Kraków, Poland}
\affiliation{Instytut Fizyki Teoretycznej, Uniwersytet Jagiello\'{n}ski, 
ulica Profesora Stanislawa Lojasiewicza 11, PL-30-348 Krak{ó}w, Poland}
\author{Krzysztof Wohlfeld}
\affiliation{Institute of Theoretical Physics, Faculty of Physics,
University of Warsaw, Pasteura 5, PL-02093 Warsaw, Poland}

\begin{abstract}
Distinguishing altermagnetism from conventional ferromagnetism and
antiferromagnetism typically relies on momentum-space probes. Here, we
show that the real-space spreading of a localized spin excitation
provides a distinctive dynamical fingerprint of altermagnetic order.
Using linear spin-wave theory on a two-dimensional checkerboard lattice
with nearest-neighbor and next-nearest-neighbor couplings $J_1$ and
$J_2$, we demonstrate that finite $J_2$ produces direction-dependent
magnon group velocities and splits the two magnon branches. The
resulting propagation remains closer to that of an antiferromagnet than
a ferromagnet, retaining an approximately circular outer wavefront,
while the direction-dependent magnon velocities produce a pronounced
cross-like spatial structure inside this front. We further show that
changing the sign of $J_2$ to enter the unfrustrated regime ($J_2<0$)
increases the characteristic propagation velocities and interchanges
the diagonal directions of enhanced propagation, corresponding to a
$\pi/2$ rotation of the anisotropic spatial pattern. These results
establish spacetime dynamics as a complementary probe of altermagnetic
magnons, providing a route to identifying unconventional magnetic order
through real-space propagation patterns in solid-state and synthetic
quantum systems.
\end{abstract}

\date{\today}

\maketitle

\section{Introduction}

For decades, the study of magnetic materials was defined by a neat dichotomy between ferromagnetism, with its net macroscopic magnetization, and antiferromagnetism, characterized by vanishing net moments and spin-degenerate bands. The recent recognition of altermagnetism has drawn intense interest by establishing a third class of collinear magnetic order---one that combines zero net magnetization with robust spin-splitting~\cite{Smejkal2022, Smejkal2022b}.

The hallmark signature of altermagnetism is its non-relativistic electronic spin splitting, which can be directly observed in reciprocal space using angle-resolved photoemission spectroscopy (ARPES)~\cite{Krempasky2024, Fedchenko2024}. However, for correlated magnets---where conventional ARPES signatures of altermagnetism are fundamentally more challenging to access~\cite{Daghofer2026}---and more generally from the perspective that ordered phases are characterized by their collective excitations, a natural question is how collective spin excitations in altermagnets differ from those in ferromagnets and antiferromagnets. Indeed, altermagnetic magnons inherit the underlying crystalline symmetries, acquiring distinct chiral character and momentum-dependent band splittings, as recently demonstrated experimentally in candidate materials such as $\alpha$-MnTe and CrSb using inelastic neutron scattering (INS) and resonant inelastic x-ray scattering (RIXS)~\cite{Liu2024, Biniskos2025}. Yet, resolving these altermagnetic magnon splittings with INS can be experimentally challenging, particularly since the energy and momentum separations of the split chiral magnons are small. In fact, in typical altermagnets they are orders of magnitude smaller than the electronic spin splittings in ARPES.

While such momentum-dependent features naturally make reciprocal-space techniques the standard probe, these experimental challenges motivate complementary approaches. More broadly, the study of quantum many-body systems has increasingly moved toward observing and characterizing dynamics directly in the time domain. Pump-probe and ultrafast spectroscopies have made it possible to follow the evolution of electronic and magnetic excitations following a controlled perturbation~\cite{Kirilyuk2010}, while quantum gas microscopes provide a complementary route to resolving and tracking individual spins directly in real space~\cite{Fukuhara2013, Gross2017}. In parallel, recent theoretical, numerical, and experimental advances have demonstrated that real-space dynamics can reveal nontrivial quantum transport and correlation phenomena, while becoming increasingly accessible through modern computational and quantum-simulation techniques~\cite{Scheie2022, Scheie2024, Kish2025, Kumaran2025, Lee2025, Rosenberg2024, Liang2025, Manovitz2025, Chowdhury2025, Haghshenas2025}. In conventional magnetic systems, such dynamics can produce characteristic real-space propagation patterns, ranging from quantum-walk behavior to ``stone-in-a-pond'' waves~\cite{Wrzosek2020}, highlighting how real-space dynamics can reveal structure that is not readily apparent in momentum-frequency space.

Importantly, the transition from momentum-frequency to real-space and real-time information does not necessarily require a time-resolved experiment. The dynamical structure factor $S(\mathbf{q},\omega)$ and its real-space, real-time counterpart are related by a Fourier transform, so conventional spectroscopic data can in principle be reorganized into a spacetime correlation function $S(\mathbf{r},t)$. This change of representation can be more than cosmetic: structures that appear as broad continua or overlapping features in $\mathbf{q}$--$\omega$ can acquire a much more organized spatial and temporal form after the transformation. Recent work on the triangular antiferromagnet $\text{KYbSe}_2$ provides a striking example~\cite{Scheie2026}. By Fourier transforming INS data to obtain $S(\mathbf{r},t)$, the authors observed a nonlinear light cone of spin correlations with dynamics intermediate between ballistic and diffusive spreading. The resulting behavior was not reproduced by several conventional theoretical descriptions and was discussed in connection with exotic collective dynamics in the vicinity of a proximate quantum-spin-liquid regime~\cite{Scheie2024}.

This example further confirms a broader principle: the real-space, real-time representation of a magnetic spectrum can expose dynamical organization that is difficult to recognize in momentum-frequency space alone. Therefore,
here we apply this perspective to altermagnetism by investigating the spacetime dynamics of localized spin excitations on a two-dimensional checkerboard Heisenberg lattice, where altermagnetic order arises from competing nearest-neighbor ($J_1$) and next-nearest-neighbor ($J_2$) exchange couplings~\cite{Yershov2024, Das2024, Kaushal2025}, see Fig.~\ref{fig:1}(a). A finite $J_2$ lifts the magnon degeneracy away from symmetry-protected nodal lines and produces momentum-dependent differences in the group velocities of the split branches. By tracking the real-space spreading of an initial single spin flip, we show that while the overall propagation boundary retains the circular footprint of the parent antiferromagnet, the directional velocity mismatch generates a prominent cross-shaped internal interference pattern. Furthermore, we demonstrate that these dynamical signatures depend critically on the sign and magnitude of $J_2$, driving a rotation of high-velocity features and changing the characteristic propagation velocities.

The remainder of the paper is organized as follows. In Sec.~\ref{sec:model}, we introduce the 2D spin-$1/2$ Heisenberg model on the checkerboard lattice, incorporating both nearest-neighbor ($J_1$) and next-nearest-neighbor ($J_2$) interactions. Using the Holstein-Primakoff transformation and linear spin-wave theory, we derive the altermagnetic magnon dispersion relations and establish the framework for calculating the time-dependent density profile of a single-spin-flip excitation. In Sec.~\ref{sec:results}, we present our central findings, demonstrating how the distinct spacetime propagation of localized spin excitations provides a dynamical probe of the underlying altermagnetic magnon structure.
Specifically, we show that the propagation retains the overall character of the parent antiferromagnet, while the altermagnetic magnon splitting produces a direction- and sublattice-dependent spatial structure whose orientation and propagation velocity depend on the sign and magnitude of $J_2$.
Finally, in Sec.~\ref{sec:conclusion}, we summarize our results and discuss their broader implications. 
The paper is supplemented by two short appendices.

\section{Model \label{sec:model}}
\begin{figure}[t]
    \centering
    \includegraphics[width=1\linewidth]{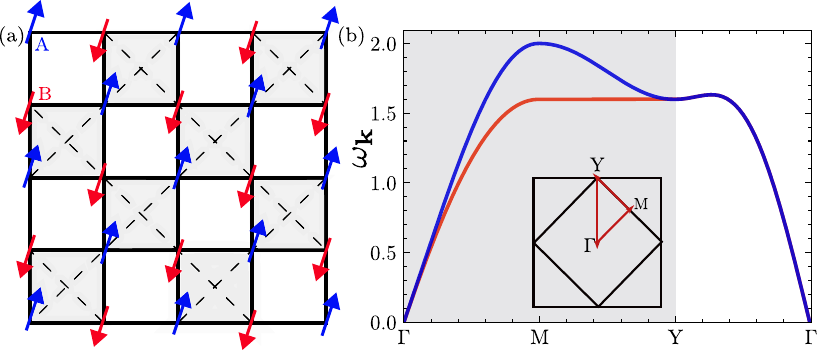}
    \caption{
    (a) Schematic representation of the Heisenberg model on a 2D checkerboard lattice featuring collinear Ne\'el order. The A and B sublattices are connected by a composite symmetry operation consisting of a mirror reflection and a spin inversion. (b) Magnon dispersion relations $\omega_{\bf k}$ for the next-nearest-neighbor Heisenberg exchange $J_2=0.2$ (with the nearest neighbor exchange $J_1=1$). The introduction of next-nearest-neighbor coupling $J_2$ unequivocally lifts the band degeneracy across the Brillouin zone, except along the high-symmetry nodal axes ($k_x=0,k_y=0$) where the degeneracy is protected by the lattice symmetry.}
    \label{fig:1}
\end{figure}
We consider a two-dimensional Heisenberg model with $S=1/2$ spin degrees of freedom:
\begin{align}
\label{eq:h}
\mathcal{H}
= J_1 \sum_{\langle \mathbf{i},\mathbf{j} \rangle}
\mathbf{S}_{\mathbf{i}}\cdot\mathbf{S}_{\mathbf{j}}
+ J_2 \sum_{\langle\langle \mathbf{i},\mathbf{j}\rangle\rangle}
\mathbf{S}_{\mathbf{i}}\cdot\mathbf{S}_{\mathbf{j}}.
\end{align}
Here, $J_1$ and $J_2$ denote the nearest- and next-nearest-neighbor interactions, respectively, on the checkerboard lattice; see Fig.~\ref{fig:1}(a).

Since the main goal of this work is to investigate the altermagnetically ordered state, we restrict our analysis to relatively small values of $|J_2|/|J_1|$. In this regime, the ground state is magnetically ordered~\cite{Canals2002, Li2015} and for nonzero $J_2$ realizes a canonical altermagnet~\cite{Yershov2024, Das2024, Kaushal2025}.
The existence of an ordered ground state allows the low-lying excited states of interest to be described in terms of noninteracting magnons within linear spin-wave theory~\cite{Auerbach1994}.

We therefore introduce Holstein--Primakoff bosons~\cite{holstein1940field, Auerbach1994}, denoted by $a$ and $b$ for the $A$ and $B$ sublattices, respectively:
\begin{equation}
\begin{aligned}
S^z_{\mathbf{i},A}
&= S-a^\dagger_{\mathbf{i}}a_{\mathbf{i}},
&\qquad
S^+_{\mathbf{i},A}
&= \sqrt{2S-a^\dagger_{\mathbf{i}}a_{\mathbf{i}}}\,a_{\mathbf{i}}
= (S^-_{\mathbf{i},A})^\dagger,\\
S^z_{\mathbf{i},B}
&= b^\dagger_{\mathbf{i}}b_{\mathbf{i}}-S,
&\qquad
S^+_{\mathbf{i},B}
&= b^\dagger_{\mathbf{i}}\sqrt{2S-b^\dagger_{\mathbf{i}}b_{\mathbf{i}}}
= (S^-_{\mathbf{i},B})^\dagger.
\end{aligned}
\end{equation}
We then expand the Holstein--Primakoff transformation in powers of $1/\sqrt{2S}$ and retain terms in the Hamiltonian up to quadratic order in the bosonic operators, corresponding to the linear spin-wave approximation (see Appendix~\ref{sec:app}).

Then applying a spatial Fourier transform over the $N_c$ unit cells (with the total number of sites $N_s = 2N_c$), $a_{\bf j} = (N_c)^{-1/2}\sum_{\mathbf{k}} e^{-i\mathbf{k}\cdot {\bf j}} a_{\mathbf{k}}$ (where $ \mathbf{k}$ is the lattice quasi-momentum vector), we can express the Hamiltonian in momentum space as
\begin{equation}
H = 
\sum_{\mathbf{k}}\Psi^\dagger_{\mathbf{k}} H_{\mathbf{k}} \Psi_{\mathbf{k}} + {\rm const.},
\end{equation}
where $\Psi^\dagger_{\mathbf{k}} = \begin{pmatrix} a^\dagger_{\mathbf{k}} & b_{-\mathbf{k}} \end{pmatrix}$. The bosonic Bogoliubov-deGennes Hamiltonian matrix $H_{\mathbf{k}}$ mixes the $\pm \mathbf{k}$ modes and takes the form:
\begin{equation}
H_{\mathbf{k}} = \begin{pmatrix}
    A_\mbf{k} & C_\mbf{k} \\
    C^*_\mbf{k} & B_\mbf{k}
\end{pmatrix}
\end{equation}
Where the coefficients 
$A_{\mathbf{k}} = 4J_1S(1-2\eta \Gamma_{\mathbf{k}1}),\,
B_{\mathbf{k}} = 4J_1S(1-2\eta \Gamma_{\mathbf{k}2}),\,\mathrm{and}\, 
C_{\mathbf{k}} = 4J_1S\gamma_{\mathbf{k}}$ depend on the lattice geometry, and we define the dimensionless ratio $\eta=J_2/4J_1$, alongside the structure factors $\Gamma_{\mathbf{k}1(2)} = 1-\cos(k_x\pm k_y)$ and $\gamma_{\mathbf{k}} = [\cos(k_x) + \cos(k_y)]/2$. 

This momentum-space Hamiltonian is diagonalized via a Bogoliubov transformation, yielding the diagonal form~\cite{Yershov2024, Kaushal2025, PhysRevLett.134.196701}:
\begin{align}
    H = \sum_{\mathbf{k},\sigma} \omega_{\mathbf{k}\sigma}
    \left(
    \alpha^\dagger_{\mathbf{k}\sigma}\alpha_{\mathbf{k}\sigma}
    + \frac{1}{2}
    \right)
    + \mathrm{const.}
\end{align}
Here, $\alpha^\dagger_{\mathbf{k}\sigma}$ is the Bogoliubov magnon creation operator. The magnon bands are given by
\begin{equation}
\label{eq:magnondisp}
\omega_{\mathbf{k}\sigma}
=
\sqrt{
\left(
\frac{A_{\mathbf{k}} + B_{\mathbf{k}}}{2}
\right)^2
-
\abs{C_{\mathbf{k}}}^2
}
-
\sigma
\left(
\frac{A_{\mathbf{k}} - B_{\mathbf{k}}}{2}
\right),
\end{equation}
where $\sigma = \pm 1$. The Bogoliubov coefficients, which will be needed below, are
\begin{equation}
\begin{aligned}
u_{\mathbf{k}}
&= \cosh \left[
-\frac{1}{2} \tanh^{-1} \left(
\frac{2\abs{C_{\mathbf{k}}}}{A_{\mathbf{k}} + B_{\mathbf{k}}}
\right)
\right], \\
v_{\mathbf{k}}
&= \sinh \left[
-\frac{1}{2} \tanh^{-1} \left(
\frac{2\abs{C_{\mathbf{k}}}}{A_{\mathbf{k}} + B_{\mathbf{k}}}
\right)
\right].
\end{aligned}
\end{equation}
Further details of the Bogoliubov transformation are provided in Appendix~\ref{sec:app}.

Although the magnon bands $\omega_{\mathbf{k}\sigma}$  are perfectly degenerate in the purely nearest-neighbor limit ($J_2 = 0$), the introduction of a finite $J_2 \neq 0$ lifts this degeneracy. As depicted in Fig.~\ref{fig:1}b, the band splitting is observed across the Brillouin zone, save for the nodal lines $k_x = 0$ and $k_y = 0$. Along these high-symmetry axes, the degeneracy is  protected by the underlying symmetries of both the Hamiltonian $H$ and the ground state—specifically, their invariance under a composite transformation that couples spin inversion with a mirror reflection interchanging sublattices $A$ and $B$~\cite{10.21468/SciPostPhys.18.4.125}. Furthermore, the branch index $\sigma$ in the dispersion relation $\omega_{\mathbf{k}\sigma}$ serves as a well-defined quantum number. It indicates the net change in the macroscopic magnetization induced by the excitation of a single magnon, yielding $\Delta S^z_{\text{tot}} = \sigma$~\cite{PhysRevLett.134.196701}.

With the model diagonalized, we now examine the real-time evolution of the system following a single spin-flip excitation at a specific site $r_o$, applied to the Hamiltonian ground state, i.e. the Bogoliubov magnon vacuum $\ket{\varnothing}$. We define the time-evolved state as 
\begin{align}
  \ket{\phi({\bf r},t)} = c_{{\bf r}} e^{-iHt/\hbar} c^\dagger_{{\bf r}_o} \ket{\varnothing},   
\end{align}
where the annihilation operator $ c_{{\bf r}}$ corresponds to either $a_{{\bf r}}$ or $b_{{\bf r}}$ depending on the parity of site ${{\bf r}}$. This leads to four distinct scenarios based on the sublattice locations of $\mathbf{r}$ and $\mathbf{r_o}$. Explicitly, when ${\bf r}_o$ lies in the A sublattice, for $\mathbf{r}\in A$, we have
\begin{equation}
\begin{aligned}
\ket{\phi({\bf r},t)} &= \frac{1}{N_c} \sum_{\mathbf{k}} e^{i (\mathbf{k}\cdot \mathbf{r} - \omega_{\mathbf{k},-} t) } u^2_{\mathbf{k}} \ket{\varnothing} \\
&\quad + \frac{1}{N_c}\sum_{\mathbf{k}\mathbf{k}'} e^{i (\mathbf{k}'\cdot \mathbf{r_i} -\mathbf{k}\cdot \mathbf{r_o} - \omega_{\mathbf{k},-} t) } \\
&\quad \times u_{\mathbf{k}} v_{\mathbf{k}'}\ket{\alpha_{\mathbf{k},-},\alpha_{-\mathbf{k}',+}}
\end{aligned}
\end{equation}
and for ${\bf r} \in B$, we have
\begin{equation}
\begin{aligned}
\ket{\phi({\bf r},t)} &= \frac{1}{N_c} \sum_{\mathbf{k}\mathbf{k}'} e^{i (\mathbf{k}'\cdot \mathbf{r} - \mathbf{k}\cdot \mathbf{r_o} - \omega_{\mathbf{k},-} t) } \\
&\quad \times \sqrt{1+\delta_{\mathbf{k},-\mathbf{k}'}} u_{\mathbf{k}} v_{\mathbf{k}'}\ket{\alpha_{\mathbf{k},-},\alpha_{-\mathbf{k}',-}}
\end{aligned}
\end{equation}
with similar expressions for $\mathbf{r_o}\in B$ given in Appendix~\ref{sec:appB}.

Finally, the main goal of the paper is to track the observable probability densities, i.e. the density profile $\rho$ defined as
\begin{align}
    \rho (\mathbf{r},t) =  \langle \phi(\mathbf{r},t) | \phi(\mathbf{r},t) \rangle. 
\end{align}
Then, we evaluate that for an initial excitation on the $A$ sublattice ($\mathbf{r_o} \in A$) the density profile is given by:
\begin{equation}
\begin{aligned}
\rho_A({\bf r},t) &= \frac{1}{N^2_c} \left|\sum_{\mathbf{k}} e^{i(\mathbf{k}\mathbf{r}-\omega_{\mathbf{k},-}t)} u^2_{\mathbf{k}}\right|^2 \\
&\quad + \frac{2}{N^2_c}\left(\sum_{\mathbf{k}} u^2_{\mathbf{k}}\right)\left(\sum_{\mathbf{k}} v^2_{\mathbf{k}}\right) \\
&\quad + \frac{2}{N^2_c}\left(\sum_{\mathbf{k}} u^2_{\mathbf{k}}v^2_{\mathbf{k}}\right) \\
&\quad + \frac{1}{N^2_c}\left|\sum_{\mathbf{k}} e^{i(\mathbf{k}\mathbf{r}-\omega_{\mathbf{k},-}t)} u_{\mathbf{k}} v_{\mathbf{k}}\right|^2.
\end{aligned}
\label{density_11}
\end{equation}
Similar expressions for $\mathbf{r_o} \in B$, i.e. for $\rho_B({\bf r},t) $, are given in  Appendix \ref{sec:appB}.

\section{Results \label{sec:results}}

\begin{figure}[t]
    \centering
    \includegraphics[width=0.9\linewidth]{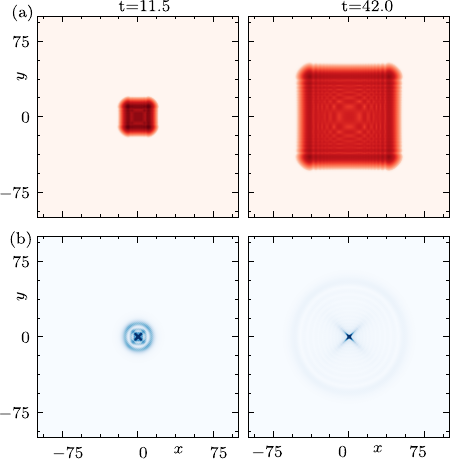}
    \caption{
Comparison of the real-space spreading of a single-site spin-flip excitation
in the absence of next-nearest-neighbor interactions ($J_2=0$).
(a) In the ferromagnetic case ($J_1<0$), the largest group velocities occur
at finite momenta, making the propagation sensitive to short-wavelength,
lattice-scale physics and producing the characteristic square-shaped density
profile.
(b) In the antiferromagnetic case ($J_1>0$), the largest group velocities
occur in the long-wavelength regime, where the linear magnon dispersion is
approximately rotationally symmetric. The resulting wavefront is therefore
nearly circular and much less sensitive to the underlying square-lattice
structure.
    }
    \label{fig:2}
\end{figure}

\begin{figure}[t]
    \centering
    \includegraphics[width=1\linewidth]{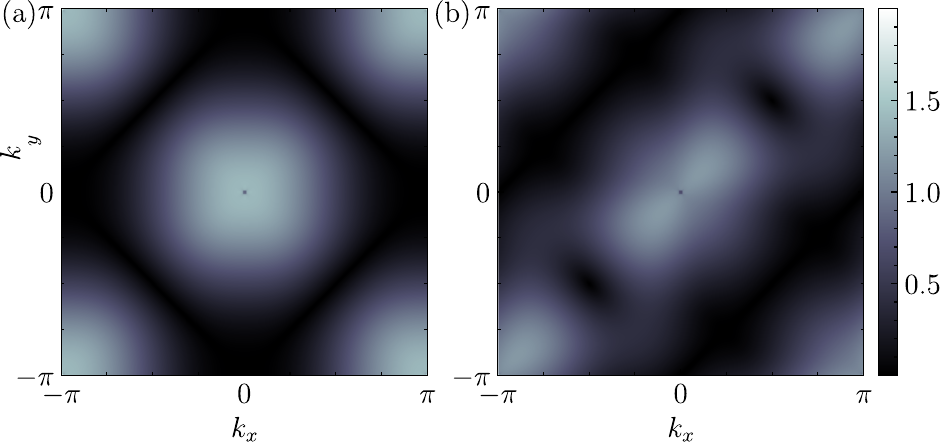}
    \caption{
Magnitude of the magnon group velocity,
$|\mathbf{v}_{\sigma}(\mathbf{k})|$, across the Brillouin zone.
(a) In the antiferromagnetic limit, $J_2=0$, the largest group velocities
are distributed nearly isotropically around the $\Gamma$ point.
(b) For finite $J_2>0$, this distribution becomes direction dependent,
with the largest velocities concentrated along selected diagonal directions.
This deformation of the group-velocity distribution reflects the
direction dependence of the altermagnetic magnon dispersion discussed in
the main text. The parameters are (a) $\theta=0$ ($J_1=1$, $J_2=0$) and
(b) $\theta=0.3$ ($J_1\simeq0.95$, $J_2\simeq0.29$), with $\sigma=-1$.
 }
    \label{fig:3}
\end{figure}
\begin{figure*}[t]
    \centering
    \includegraphics[width=1\linewidth]{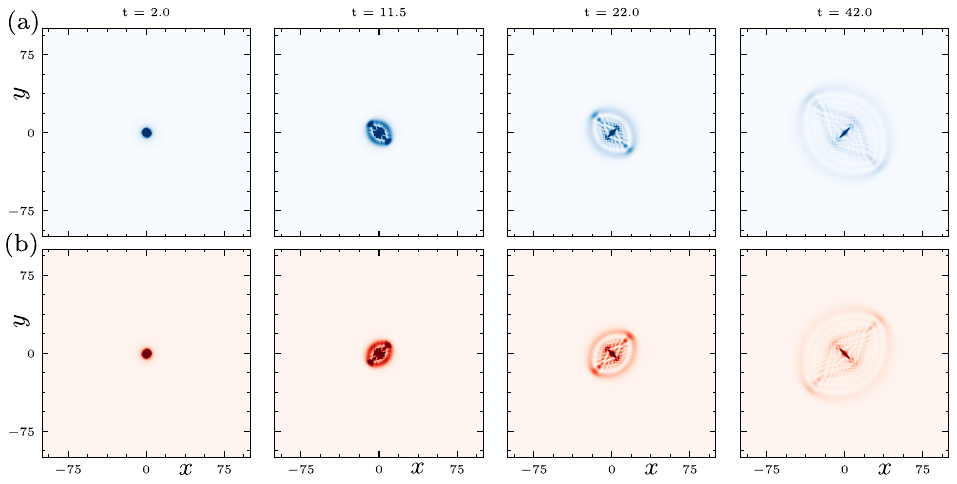}
    \caption{
Time-resolved spatial density profiles following a single-site spin flip in
the representative altermagnetic regime, $\theta=0.3$ ($J_1\simeq0.95$, $J_2\simeq0.29$), at
$t=2.0,\,11.5,\,22.0$, and $42.0$. The outer propagation front remains
approximately circular, reflecting the dominant antiferromagnetic-like
propagation, while the direction-dependent base dispersion and the
altermagnetic branch splitting produce a cross-like structure inside the
wave packet. Panels (a) and (b) show the $\sigma=+1$ and $\sigma=-1$
magnon branches, respectively. The lattice size is $N=200$.}
    \label{fig:4}
\end{figure*}
\begin{figure*}[t]
    \centering
    \includegraphics[width=1\linewidth]{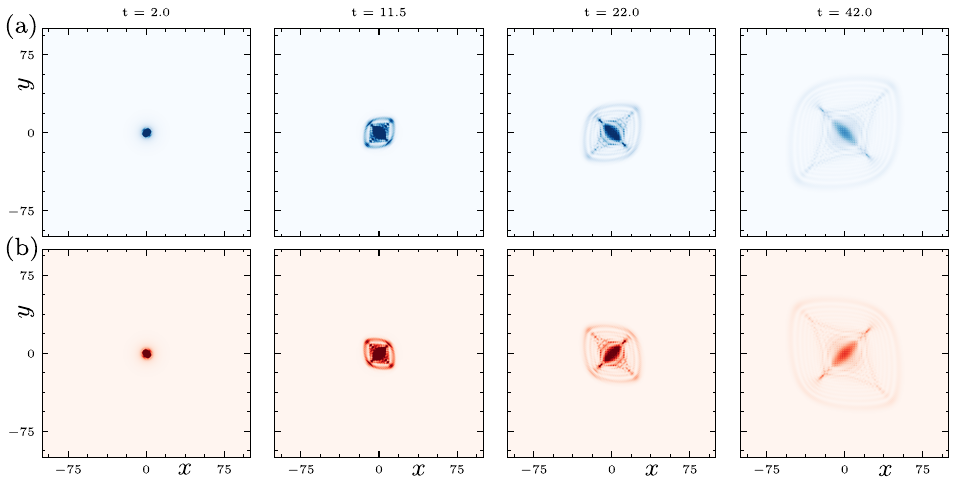}
    \caption{
Time-resolved spatial density profiles following a single-site spin flip
in the unfrustrated altermagnetic regime, $\theta=-0.4$
($J_1\simeq0.92$, $J_2\simeq-0.38$), at
$t=2.0,\,11.5,\,22.0$, and $42.0$. Compared with the frustrated regime
shown in Fig.~\ref{fig:4}, the wave packet spreads considerably farther
and its anisotropic spatial pattern is rotated by $\pi/2$. These changes
reflect the larger characteristic magnon group velocities for $J_2<0$
and the reversal of the branch-dependent altermagnetic contribution
discussed in the main text. Panels (a) and (b) show the $\sigma=+1$
and $\sigma=-1$ magnon branches, respectively. The lattice size is
$N=200$.
}
    \label{fig:5}
\end{figure*}

\subsection{Ferromagnetic and antiferromagnetic limits}

We start by revisiting the simplest case in which next-nearest-neighbor
interactions are absent ($J_2=0$) and the spin exchange is ferromagnetic
($J_1<0$), see~\cite{Wrzosek2020}. The ground state is then fully polarized
and acts as a true vacuum for magnons, with no quantum fluctuations. The
magnon excitations are characterized by the well-known ferromagnetic
dispersion
$ \omega_{\mathbf{k}} = 4S|J_1|(1-\gamma_{\mathbf{k}})$,
which is quadratic in momentum, $\omega_{\mathbf{k}}\propto k^2$, near the
center of the Brillouin zone (the $\Gamma$ point). Because the highest magnon
group velocities originate from regions of the Brillouin zone away from its
center, the spatial spreading of a localized spin flip is dominated by large
momenta and thus by short-wavelength, lattice-scale physics that strongly
reflects the underlying lattice symmetry~\cite{Wrzosek2020}.
Consequently, the dynamics following a single spin flip map onto a
continuous-time ``quantum walk'' of a single particle, producing the
characteristic square-shaped spatial density profile shown in
Fig.~\ref{fig:2}(a).

The situation is qualitatively different in the antiferromagnet ($J_1>0$ and
$J_2=0$), whose ordered ground state is dressed by quantum fluctuations. The
corresponding magnon dispersion is
$\omega_{\mathbf{k}}
    = 4SJ_1\sqrt{1-\gamma_{\mathbf{k}}^2}$.
Near the $\Gamma$ point, it becomes linear in momentum,
$\omega_{\mathbf{k}}\propto|\mathbf{k}|$, and hence the maximum group velocity is
reached in this long-wavelength limit. The spreading of a localized spin flip
is therefore dominated by long-wavelength excitations and resembles the
propagation of a ``stone-in-a-pond'' wave~\cite{Wrzosek2020}. Since the
long-wavelength dispersion is approximately rotationally symmetric, the
resulting wavefront is nearly circular and much less sensitive to the
underlying square-lattice structure [see Fig.~\ref{fig:2}(b)].

\subsection{Altermagnetic magnon propagation}

Having established the ferromagnetic and antiferromagnetic limits, we now
consider the altermagnetic regime obtained for finite $J_2$. We parameterize
the couplings as $J_1=\cos\theta$ and $J_2=\sin\theta$, keeping their overall
scale fixed. We first focus on a representative altermagnetic case obtained at $\theta=0.3$,
corresponding to dominant $J_1\simeq0.95$ and somewhat weaker $J_2\simeq0.29$.

A finite $J_2$ modifies the magnon dispersion in two related ways. First, it
changes the part of the dispersion common to both magnon branches. In the
antiferromagnetic limit $J_2=0$, the long-wavelength dispersion around the
$\Gamma$ point is approximately rotationally symmetric (despite the
underlying square lattice), depending to leading order only on
$|\mathbf{k}|$. For $J_2\neq0$, the next-nearest-neighbor interactions
introduce an additional dependence on the lattice diagonals through
$\cos(k_x+k_y)$ and $\cos(k_x-k_y)$. This deforms the nearly isotropic
antiferromagnetic dispersion and makes its slope direction dependent. We
refer to this common part of the two magnon branches as the \emph{base
dispersion}. Since the magnon group velocity is
\begin{equation}
    \mathbf{v}_{\sigma}(\mathbf{k})
    = \nabla_{\mathbf{k}}\omega_{\mathbf{k}\sigma},
\end{equation}
this deformation directly produces a direction-dependent contribution to
the propagation velocity.

Second, the diagonal $J_2$ couplings enter differently for the two
sublattices and lift the degeneracy of the two magnon branches away from the
symmetry-protected nodal lines. This altermagnetic splitting gives an
additional, branch-dependent contribution to the group velocity. Thus, the
dynamics for finite $J_2$ reflect two effects: the directional dependence of
the base dispersion and the difference between the two split magnon
branches.

Figure~\ref{fig:3} illustrates these effects explicitly in the calculated group velocities.
As discussed in Sec.~III\,A, for $J_2=0$ the distribution of the largest
group velocities is nearly isotropic around the $\Gamma$ point. For finite
$J_2$, this distribution is deformed, with the largest velocities
concentrated along selected diagonal directions. In addition, the two magnon
branches acquire different velocities away from the nodal lines.

The consequences are directly visible in spacetime magnon dynamics. A spin flip localized
on a single lattice site contains all momenta, and its time evolution
therefore probes the full magnon dispersion. Figure~\ref{fig:4}
shows the resulting density profile for $\theta =0.3$. Since $J_1$ remains the
dominant coupling, the outer propagation front remains approximately
circular, as in the parent antiferromagnet. Inside this front, however, the
direction-dependent base velocity together with the altermagnetic branch
splitting produces the cross-like structure visible in the density profile.
The coexistence of a nearly circular outer front with a structured interior
is therefore a characteristic real-space consequence of the altermagnetic
magnon dispersion.

Overall, the propagation in the altermagnets remains much
closer to that of the parent antiferromagnet than to the ferromagnetic case
discussed in Sec.~III\,A. In particular, the outer wavefront remains
approximately circular, reflecting the dominant long-wavelength
antiferromagnetic propagation. The main effect of altermagnetism is therefore
not to replace this behavior, but to superimpose a pronounced
direction-dependent structure inside the wave packet. In this limited sense,
the internal pattern is reminiscent of the ferromagnetic case, where
finite-momentum modes also make the propagation sensitive to the underlying
lattice. The crucial difference is that in the altermagnet this
lattice-sensitive structure appears within an otherwise
antiferromagnetic-like expanding wavefront.

\begin{figure}[t]
    \centering
    \includegraphics[width=0.9\linewidth]{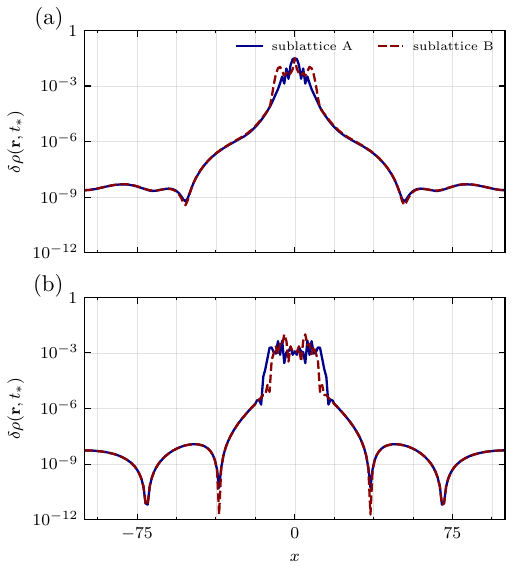}
    \caption{
Sublattice-resolved density profiles at $t=11.5$ along the diagonal
$y=x$ for (a) the frustrated regime, $\theta=0.3$, and (b) the
unfrustrated regime, $\theta=-0.3$. Reversing the sign of $J_2$
interchanges the enhanced propagation between the A and B sublattices
along this diagonal. The spatial distribution is also broader for
$J_2<0$, reflecting the larger characteristic propagation velocity in
the unfrustrated regime.
}
    \label{fig:6}
\end{figure}
\begin{figure*}[t]
    \centering
    \includegraphics[width=1\linewidth]{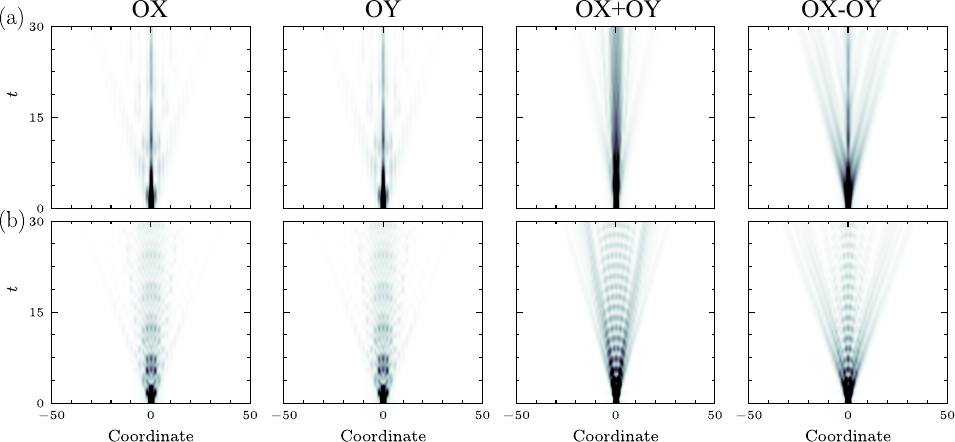}
    \caption{
Spatiotemporal density profiles following a single-site spin flip along
the principal (OX, OY) and diagonal (OX+OY, OX-OY) lattice directions.
(a) Frustrated regime, $\theta=0.3$ ($J_1\simeq0.96$, $J_2\simeq0.30$),
for the $\sigma=+1$ magnon branch.
(b) Unfrustrated regime, $\theta=-0.4$ ($J_1\simeq0.92$,
$J_2\simeq-0.39$), for the $\sigma=-1$ magnon branch.
The broader propagation cones in (b) illustrate the larger
characteristic magnon velocities in the unfrustrated regime.
The lattice size is $N=200$.
}
    \label{fig:7}
\end{figure*}

\subsection{Role of frustration}

We next examine how the sign of $J_2$ affects altermagnetic magnon
propagation by comparing the frustrated ($J_2>0$) and unfrustrated
($J_2<0$) regimes. Figure~\ref{fig:5} shows two main changes that encountered once $J_2$ is no longer frustrating $J_1$. First,
the wave packet spreads considerably farther in the unfrustrated regime,
indicating larger characteristic propagation velocities. Second, the
directional structure of the wave packet is interchanged between the two
lattice diagonals, corresponding to a $\pi/2$ rotation of the
anisotropic spatial pattern.

The sublattice dependence of the latter effect is resolved more clearly
in Fig.~\ref{fig:6}, which shows the density at $t=11.5$ along the
diagonal $y=x$. For $J_2>0$, the wave packet extends farther along this
diagonal on sublattice A than on sublattice B, whereas reversing the
sign of $J_2$ interchanges this behavior. Thus, changing the sign of
$J_2$ reverses which sublattice exhibits the enhanced propagation along
a given diagonal. Figure~\ref{fig:6} also shows that the overall spatial
distribution is broader for $J_2<0$.

The difference in propagation speed is also nicely seen in the so-called light cones. Figure~\ref{fig:7} shows the ``continuous'' time evolution of the magnon density profile along the principal
and diagonal lattice directions. The different slopes of the
propagating fronts demonstrate that the broader spatial profiles are
associated with different propagation velocities rather than only with
a redistribution of spectral weight. In particular, the unfrustrated
case displays a substantially wider propagation cone. 

Taken together,
Figs.~\ref{fig:5}--\ref{fig:7} therefore reveal two effects of changing
the sign of $J_2$: a change in the characteristic propagation speed and
a reversal of the directional, sublattice-dependent structure.
The origin of these two effects can be understood directly from the
magnon dispersion. Equation~\eqref{eq:magnondisp} can be written as
\begin{equation}
\omega_{\mathbf{k}\sigma}
=
\omega_{\rm base}(\mathbf{k})
+
\sigma S J_2
\left(
\Gamma_{\mathbf{k}1}-\Gamma_{\mathbf{k}2}
\right),
\label{eq:omega_decomposition}
\end{equation}
where
\begin{equation}
\omega_{\rm base}(\mathbf{k})
=
\sqrt{
\left[
4J_1S-SJ_2
\left(
\Gamma_{\mathbf{k}1}+\Gamma_{\mathbf{k}2}
\right)
\right]^2
-
|C_{\mathbf{k}}|^2
}.
\label{eq:omega_base}
\end{equation}
The group velocity can therefore be separated into
\begin{equation}
\mathbf{v}_{\sigma}(\mathbf{k})
=
\mathbf{v}_{\rm base}(\mathbf{k})
+
\mathbf{v}_{{\rm alt},\sigma}(\mathbf{k}),
\label{eq:v_decomposition}
\end{equation}
where the two terms account for the overall propagation scale and the
branch-dependent directional structure, respectively.

We first consider the base velocity. Near the $\Gamma$ point,
\begin{equation}
\Gamma_{\mathbf{k}1}+\Gamma_{\mathbf{k}2}
\simeq k^2,
\qquad
\gamma_{\mathbf{k}}
\simeq
1-\frac{k^2}{4},
\end{equation}
so that
\begin{equation}
\omega_{\rm base}(\mathbf{k})
\simeq
S\sqrt{8J_1(J_1-J_2)}\,|\mathbf{k}|.
\label{eq:omega_base_lw}
\end{equation}
The corresponding long-wavelength group velocity is
\begin{equation}
v_{\rm base}
\simeq
S\sqrt{8J_1(J_1-J_2)}.
\label{eq:v_base_lw}
\end{equation}
For $J_1>0$, this velocity increases when $J_2<0$ and decreases when
$J_2>0$. Thus, the unfrustrated next-nearest-neighbor interaction
enhances the long-wavelength antiferromagnetic-like propagation,
whereas frustration suppresses it. Although intuitively such result could have been expected, this provides a direct explanation
for the faster spreading observed in the unfrustrated regime. 

The directional and sublattice-dependent behavior originates from the
second term in Eq.~\eqref{eq:omega_decomposition}. Using
\begin{equation}
\Gamma_{\mathbf{k}1}-\Gamma_{\mathbf{k}2}
=
2\sin k_x\sin k_y,
\end{equation}
its contribution to the group velocity is
\begin{equation}
\mathbf{v}_{{\rm alt},\sigma}(\mathbf{k})
=
2\sigma S J_2
\left(
\cos k_x\sin k_y,\,
\sin k_x\cos k_y
\right).
\label{eq:v_alt}
\end{equation}
This contribution changes sign under either
$J_2\rightarrow-J_2$ or $\sigma\rightarrow-\sigma$. Reversing $J_2$
therefore interchanges the branch-dependent velocities. Since the
sublattice-resolved dynamics are tied to opposite magnon branches, this
accounts for the interchange between A and B observed in
Fig.~\ref{fig:6}.

Finally, the same term explains the rotation of the directional pattern.
Under a $\pi/2$ rotation,
$(k_x,k_y)\rightarrow(-k_y,k_x)$,
\begin{equation}
\sin k_x\sin k_y
\rightarrow
-\sin k_x\sin k_y.
\end{equation}
A $\pi/2$ rotation therefore reverses the branch-dependent part of the
dispersion in the same way as $J_2\rightarrow-J_2$. This explains why
reversing the sign of $J_2$ interchanges the two diagonal directions of
enhanced propagation seen in Fig.~\ref{fig:5}.

\section{Conclusion \label{sec:conclusion}}

\subsection{Summary}
In summary, we have theoretically characterized the spacetime dynamics
of localized spin excitations in a two-dimensional checkerboard
altermagnet and compared their propagation with the conventional
ferromagnetic and antiferromagnetic limits. Using the
Holstein--Primakoff transformation and linear spin-wave theory, we
showed that finite next-nearest-neighbor interactions ($J_2$) lift the
magnon branch degeneracy away from symmetry-protected nodal lines and
produce momentum- and branch-dependent group velocities.

These features manifest directly in the real-time spreading of a
localized spin flip. In the altermagnetic regime with $J_2>0$, the
outer propagation front remains approximately circular, retaining the
dominant long-wavelength character of the parent antiferromagnet.
Inside this front, however, the direction-dependent base dispersion
and the splitting of the two magnon branches produce a pronounced
cross-like spatial structure. The coexistence of an
antiferromagnetic-like outer wavefront with a direction-dependent
internal pattern therefore provides a characteristic real-space
signature of the altermagnetic magnon dispersion.

We further showed that the propagation depends strongly on the sign of
$J_2$. In the unfrustrated regime ($J_2<0$), the characteristic
propagation velocity is enhanced relative to the frustrated case. This
follows directly from the long-wavelength base velocity,
$v_{\rm base}\propto\sqrt{J_1(J_1-J_2)}$, which increases for $J_2<0$
and decreases for $J_2>0$. At the same time, the branch-dependent
contribution to the group velocity is proportional to $\sigma J_2$ and
therefore reverses upon changing the sign of $J_2$. This accounts for
the interchange of the sublattice-dependent propagation and the
$\pi/2$ rotation of the directional spatial pattern. Together, these
results show how both the magnon propagation speed and its directional
structure encode the underlying altermagnetic interactions.

\subsection{Experimental outlook}

The spacetime patterns identified here are, in principle, accessible to bulk spectroscopies, with the two-sublattice structure providing an additional experimental handle beyond the total dynamical structure factor. In a unit cell containing two sublattices, the INS cross section contains interference terms between basis sites, with their relative phase determined by the momentum transfer $\mathbf{q}$. Consequently, measurements at different reciprocal-lattice positions probe distinct combinations of intra- and inter-sublattice dynamical correlations. A sufficiently complete single-crystal dataset of the $(\mathbf{q},\omega)$-dependent response can therefore be used to disentangle these basis contributions and, after Fourier transformation, reconstruct sublattice-resolved real-space correlation functions. Polarization analysis can provide further separation of spin and chiral components, while the momentum dependence of the magnetic structure factor carries the essential information about the basis-site interference. Although such a reconstruction requires broader momentum coverage than a conventional Fourier transform of the total structure factor $S(\mathbf{q},\omega)$, it establishes that the sublattice-resolved spacetime dynamics considered here are, in principle, experimentally accessible in bulk solids rather than being restricted to an idealized theoretical construction.

RIXS provides a complementary route in which the sublattice sensitivity can be built directly into the scattering matrix element. Unlike neutron scattering, magnetic RIXS can have amplitudes that depend sensitively on the local orbital environment and photon polarization, allowing different components of the magnetic response to be selectively enhanced through interference between contributions from inequivalent sites~\cite{Marra2012}. This mechanism is particularly relevant to altermagnetic systems, where the magnetic sublattices are related by nontrivial crystal symmetries rather than by a simple translational operation. Recent RIXS studies of altermagnetic magnons have demonstrated sensitivity to their sublattice and chiral character~\cite{Biniskos2025}, while sublattice-specific circular dichroism has more recently been proposed and observed in the zero-moment altermagnet Fe$_2$Mo$_3$O$_8$~\cite{Channagowdra2025}. These developments suggest that, for an altermagnetic lattice such as the checkerboard model considered here, an appropriately chosen RIXS geometry and polarization could assign different weights to the two magnetic sublattices and thereby enhance the distinct magnon branches. RIXS could thus provide an experimentally complementary view of the same momentum-dependent sublattice structure that, in our calculation, produces the characteristic real-space interference pattern.

A particularly direct realization of the present proposal is offered by putting altermagnetism on a programmable quantum simulator~\cite{Das2024}. In optical-lattice and quantum-gas-microscope platforms, local spin preparation, single-site addressing, and repeated imaging can provide direct access to the propagation of a localized excitation in real space and time. This makes it possible to measure not only the overall spreading boundary but also the internal interference structure and its evolution as the exchange parameters are tuned. The combination of controlled $J_1$--$J_2$ interactions and site-resolved detection therefore provides a concrete route to testing the predicted rotation of high-velocity features across the frustrated and unfrustrated regimes. More broadly, our results show that the characteristic spacetime interference generated by momentum-dependent altermagnetic magnon velocities can serve as a direct dynamical signature of symmetry-driven magnetic order, providing a bridge between momentum-space spectroscopy in solid-state materials and real-space, real-time measurements in synthetic quantum systems.

\section*{acknowledgments}
We thank Krzysztof Sacha and Pedro M. C\^onsoli for helpful discussions.
A.E.K and K. W. thanks the National Science Center,
Poland for financial support (Grant No. 2024/55/B/ST3/
03144). K.S.E acknowledges the support of the National Science Centre, Poland~(Grant No. 2021/42/A/ST2/00017).
This research was carried out with the support of the Interdisciplinary Center for Mathematical and Computational Modeling at the University of Warsaw (ICM UW) under grant no. g105-2771 and g105-2772. The numerical computations in this work were supported in part by PL-Grid Infrastructure~(Grant No. PLG/2025/018837).
\bibliography{altmagspacetime} 

\appendix
\section{The linear spin wave Hamiltonian and the Bogoliubov transformation}
\label{sec:app}
Starting from the spin Hamiltonian in Eq.~\eqref{eq:h} of the main text and 
retaining terms up to quadratic order in the Holstein–Primakoff bosons, we obtain the linear(ised) spin-wave Hamiltonian $H$:
\begin{equation}
\begin{aligned}
H &= 2J_1 S\sum_{\langle \mathbf{i}\mathbf{j} \rangle} 
\left[ a^\dagger_{\mathbf{i}}a_{\mathbf{i}} + b^\dagger_{\mathbf{j}} b_{\mathbf{j}} 
+ a_{\mathbf{i}}b_{\mathbf{j}} + a^\dagger_{\mathbf{i}} b^\dagger_{\mathbf{j}} \right] \\
&\quad - 2 J_2S\sum_{\langle\langle \mathbf{i}\mathbf{j}\rangle\rangle} 
\left[ a^\dagger_{\mathbf{i}} a_{\mathbf{i}} - a_{\mathbf{i}} a^\dagger_{\mathbf{j}} 
- a_{\mathbf{j}} a^\dagger_{\mathbf{i}}\right] \\
&\quad - 2 J_2S\sum_{\langle\langle \mathbf{i}\mathbf{j}\rangle\rangle} 
\left[ b^\dagger_{\mathbf{i}} b_{\mathbf{i}} - b_{\mathbf{i}} b^\dagger_{\mathbf{j}} 
- b_{\mathbf{j}} b^\dagger_{\mathbf{i}}\right] \\
&\quad + N_sS^2(J_2-2J_1).
\end{aligned}
\end{equation}

Then, by introducing the Fourier transform on each sub-lattice, we obtain the ${\bf k}-$space Bogoliubov-deGennes Hamiltonian $H_{\mathbf{k}}$. The magnon branches are then obtained by solving for the eigenmodes and eigenfrequencies of $\sigma^z_{\mathbf{k}} H_{\mathbf{k}}$---yielding:
\begin{equation}
H = N_s S\,(S+1)(J_2-2J_1) + \sum_{\mathbf{k},\sigma} \omega_{\mathbf{k}\sigma}\left(\alpha^\dagger_{\mathbf{k}\sigma}\alpha_{\mathbf{k}\sigma} + \frac{1}{2}\right),
\end{equation}
where $\sigma=\pm 1$, and the corresponding magnon modes are given by:
\begin{equation}
\begin{pmatrix}
\alpha_{\mathbf{k},-} \\
\alpha_{-\mathbf{k},+}^{\dagger}
\end{pmatrix}
=
\begin{pmatrix}
u_{\mathbf{k}} & -v_{\mathbf{k}}e^{i\varphi_{\mathbf{k}}} \\
-v_{\mathbf{k}}e^{-i\varphi_{\mathbf{k}}}  & u_{\mathbf{k}}
\end{pmatrix}
\begin{pmatrix}
a_{\mathbf{k}} \\
b_{-\mathbf{k}}^{\dagger}
\end{pmatrix},
\end{equation}
with $\varphi_{\mathbf{k}}=\arg(C_{\mathbf{k}})$  and the Bogoliubov coefficients $u_{\bf k}$, $v_{\bf k}$ as well as $A_{\mathbf{k}}$, $B_{\mathbf{k}}$, $C_{\mathbf{k}}$ factors defined in the main text.

\section{Derivation of the excitation density for an excitation on the B sublattice}
\label{sec:appB}

We derive the equations governing the evolution of the system following a local excitation on the B sublattice, i.e. for ${\bf r}_0 \in B $. Then, for $\mathbf{r} \in B$, we have
\begin{equation}
\begin{aligned}
\ket{\phi(\mathbf{r},t)}
&= \frac{1}{N_c} \sum_{\mathbf{k}}
e^{-i(\mathbf{k}\cdot\mathbf{r}+\omega_{\mathbf{k},+}t)}
u^2_{\mathbf{k}}\ket{\varnothing} \\
&\quad + \frac{1}{N_c}\sum_{\mathbf{k},\mathbf{k}'}
e^{-i(\mathbf{k}'\cdot\mathbf{r}-\mathbf{k}\cdot\mathbf{r}_o
+\omega_{\mathbf{k},+}t)} \\
&\quad \times u_{\mathbf{k}}v_{\mathbf{k}'}
\ket{\alpha_{\mathbf{k},+},\alpha_{\mathbf{k}',-}}.
\end{aligned}
\end{equation}
For $\mathbf{r} \in A$, we instead have
\begin{equation}
\begin{aligned}
\ket{\phi(\mathbf{r},t)}
&= \frac{1}{N_c}\sum_{\mathbf{k},\mathbf{k}'}
e^{-i(\mathbf{k}'\cdot\mathbf{r}-\mathbf{k}\cdot\mathbf{r}_o
+\omega_{\mathbf{k},+}t)} \\
&\quad \times
\sqrt{1+\delta_{\mathbf{k},-\mathbf{k}'}}
u_{\mathbf{k}}v_{\mathbf{k}'}
\ket{\alpha_{\mathbf{k},+},\alpha_{-\mathbf{k}',+}}.
\end{aligned}
\end{equation}

The corresponding excitation density can then be written as
\begin{equation}
\begin{aligned}
\rho_B(\mathbf{r},t)
&= \frac{1}{N_c^2}
\left|
\sum_{\mathbf{k}}
e^{i(\mathbf{k}\cdot\mathbf{r}+\omega_{\mathbf{k},+}t)}
u^2_{\mathbf{k}}
\right|^2 \\
&\quad + \frac{2}{N_c^2}
\left(\sum_{\mathbf{k}}u^2_{\mathbf{k}}\right)
\left(\sum_{\mathbf{k}}v^2_{\mathbf{k}}\right) \\
&\quad + \frac{2}{N_c^2}
\left(\sum_{\mathbf{k}}u^2_{\mathbf{k}}v^2_{\mathbf{k}}\right) \\
&\quad + \frac{1}{N_c^2}
\left|
\sum_{\mathbf{k}}
e^{i(\mathbf{k}\cdot\mathbf{r}+\omega_{\mathbf{k},+}t)}
u_{\mathbf{k}}v_{\mathbf{k}}
\right|^2.
\end{aligned}
\label{density_22}
\end{equation}

\end{document}